\documentclass[twocolumn,amsmath,amssymb,showpacs,superscriptaddress,prb,aps,longbibliography,10pt]{revtex4-1}
\usepackage{graphicx}
\usepackage{bm}
\usepackage{dcolumn}
\usepackage{amssymb}
\usepackage{amsmath}
\usepackage{amsfonts}
\usepackage{epsfig}
\usepackage{graphicx}
\usepackage{stackrel}
\usepackage{paralist}
\usepackage[rgb]{xcolor}
\usepackage{todonotes}
\usepackage{pdfcomment}
\usepackage{mathrsfs}
\usepackage{subfigure}
\def\beas{\begin{eqnarray*}}
\def\eeas{\end{eqnarray*}}
\def\bea{\begin{eqnarray}}
\def\eea{\end{eqnarray}}
\def\be{\begin{equation}}
\def\ee{\end{equation}}

\newcommand{\bpm}{\begin{pmatrix}}
\newcommand{\epm}{\end{pmatrix}}
\newcommand{\bmm}{\begin{matrix}}
\newcommand{\emm}{\end{matrix}}

\begin{document}

% Page header
%\markboth{Ady Stern}{Fractional Topological Insulators}

% Title
\title{Fractional Topological Insulators - a pedagogical review}
\author{Ady Stern}
\affiliation{Department of Condensed Matter Physics, Weizmann Institute of Science, Rehovot 76100, Israel}

%Abstract
\begin{abstract}
Fractional topological insulators are electronic systems that carry fractionally charged excitations, conserve charge and are symmetric to reversal of time. In this review we introduce the basic essential concepts of the field, and then survey theoretical understanding of fractional topological insulators in two and three dimensions. In between, we discuss the case of "two and a half dimensions", the fractional topological insulators that may form on the two dimensional surface of an unfractionalized three dimensional topological insulator. We focus on electronic systems and emphasize properties of edges and surfaces, most notably the stability of gapless edge modes to perturbations.
\end{abstract}

%Keywords, etc.
%\begin{keywords}
%Topological states of matter, Fractional topological insulators, Fractionalized states
%\end{keywords}

\maketitle

%Table of Contents
\tableofcontents

% Heading 1
\section{INTRODUCTION}
Looking at it from a distance, one sees topological states of matter as a surprising outcome of condensed matter physics, where systems that are full of details give rise to physical phenomena that are astonishingly independent of these details. The most striking example, the quantum Hall effect, shows a quantization of the Hall resistivity to a level better than one part in $10^9$, with the quantum being as universal as physics gets, $h/e^2$. The discovery of the integer quantum Hall effect exposed the way that localization gives rise to precision. Shortly thereafter, the fractional quantum Hall effect exposed the way in which interaction between electrons may lead to wide spread fractionalization - of the charge, the spin, the statistics, the central charge, and of several response functions.

With this history fresh in memory, it is only natural to hope that the discovery of the time-reversal-symmetric analogs of the quantum Hall effect, the topological insulators, would be followed by a discovery of interaction-driven set of fractionalized phases that are symmetric to time reversal, dubbed the fractional topological insulators. This review attempts at a pedagogical survey of these phases. At present this is a theoretical field, since no unambiguous observation of such a phase has been reported. Furthermore, the present theory is mostly based on idealized models with no straightforward application to a well defined material. Nevertheless, as a theoretical field of study it led to several interesting outcomes which in principle have observable consequences.

The structure of the review is the following: In Sec. (\ref{basicconcepts}) we present the basic notions and shortly review some technical aspects.
Readers fluent in the physics of the fractional quantum Hall effect and in the physics of topological insulators of Bloch electrons may probably skip this Section.
In Sec. (\ref{twod}) we discuss two dimensional (2D) fractional topological insulators. In (\ref{twodonthreed}) we  go half a dimension higher and introduce the fractional topological insulators that may be formed on the two dimensional surfaces of three dimensional non-fractionalized topological insulators. In Sec. (\ref{threed}) we climb up the remaining half a dimension and review fractional topological insulators in three dimensions (3D).

Note that the review focuses on fractional topological insulators that are symmetric to time reversal, and leaves outside all other fractionalized phases.  Furthermore, we de-emphasize subjects that were emphasized in recent reviews, particularly \cite{FieteMaciejko2015} and \cite{JasonPaul}. The latter is part of this volume.

%Heading 1
\section{Basic concepts \label{basicconcepts}}
In this section we explain in some detail what we mean by fractional topological insulators and give a quick review of the few technical tools that will be used.

% Heading 2
\subsection{What do we mean by Insulators?}
Most of us have learned what insulators are at a single-digit age, when explained why not to stick a needle into the electric socket. Therefore, this subsection is here just to limit the scope of the review. We will describe electronic systems where the electrons are in principle free to move between lattice sites. Thus, when the gap closes, the system may conduct electric current. We will not deal with systems where the positions of the electrons are frozen and the only active degrees of freedom are the electrons' spins.
\subsection{What do we mean by ``Topological Insulators''?}

"Topological Insulators" \cite{BHZ,FuTI,MooreTI,RoyTI,KaneReview,QiReview,MooreReview,TopologicalFieldTheoryTI} have become a widely used term, reaching as far as "the Big Bang" (the prime time TV series, not the singular event that is rumored to have happened a long time ago). What do we mean (here at least) when we use it? We consider gapped systems, i.e., systems where the Hamiltonian $H$ displays an energy gap between the ground state and the first excited state. We specify a set of symmetries, and we then group under the same equivalence class all the Hamiltonians  $H'$ to which $H$ may be deformed without a closure of the energy gap and without a violation of the specified symmetries. Topological insulators are insulators that do not share the same class with lattices in the atomic limit, i.e., lattices where electrons cannot hop between sites.

We need to complement this definition with some comments, some examples and some consequences. The systems we think of are at the thermodynamic limit, and are positioned on a compact geometry. The introduction of an edge may lead to the occurrence of gapless edge excitations, which will be elaborated on below.  The ground state in the compact geometry may become degenerate in the thermodynamic limit. This degeneracy will be very important in what follows - it is a defining feature of a fractionalized state. In the presence of impurities the energy gap is not strictly an energy gap, but rather a mobility gap. Finally the symmetries may be time reversal symmetry, particle-hole symmetry, charge conservation, lattice symmetries etc. In this review we will almost always assume charge conservation and time reversal symmetry to hold (although we occasionally use the quantum Hall effect as a reference). The topological classification we get obviously depends on the set of symmetries we choose.

The first, and still most striking, example of a topological insulator is the quantum Hall effect \cite{QHE}. Non-interacting electrons confined to two dimensions and subjected to a perpendicular magnetic field form Landau levels, and when their density is such that exactly an integer number of Landau levels is filled, the system is gapped. At a first sight this may seem like an esoteric observation, since the lines of an integer filling factor are of measure zero in the density-magnetic field plane. At a deeper sight, however, it is very profound, since the introduction of disorder and its localizing effect on electronic states make these lines develop a width and introduces the quantum Hall plateaus.

The quantum Hall effect is an illustrative example to some fundamental concepts of topological states of matter. Of these, let us mention bulk-based topological invariants, gapless edge modes  and bulk-edge correspondence. The quantized Hall conductivity that gives the effect its name may be (theoretically) observed on a torus geometry, where a time variation of the flux through one hole of the torus induces an electric field that makes a Hall current flow around the  other hole of the torus. That current may be calculated to linear response in the electric field, and the Hall conductivity may be expressed in terms of the derivatives of the ground state wave function with respect to the fluxes within the two holes.  Under the assumption of a single ground state, the resulting Hall conductivity is an integer number of $e^2/h$, just as seen experimentally at the integer quantum Hall effect \cite{TKNN,ThoulessNiu}.  Experiments are carried out on geometries with edges, and in these geometries the conservation of charge and energy requires the existence of  chiral gapless edge modes. In fact, under certain experimental conditions the entire current is carried by these edge modes. Consequently, it is essential that the properties of the bulk and the edge will be correlated.

Although in the presence of time reversal symmetry the Hall conductivity is obviously zero, a topological classification may be carried out based on other topological invariants. Interestingly, this classification is not trivial also in three dimensions. For non-interacting electrons in the presence of time reversal symmetry and charge conservation, both in two and in three dimensions, there is a $Z_2$ classification that distinguishes trivial (i.e., systems that may be adiabatically connected to a lattice in the atomic limit) from topological systems.

%The quantized Hall conductivity of a gapped Hamiltonian is a topological invariant that cannot be changed without a closure of the energy gap. Other topological states are not as strongly protected. Rather, they are characterized by topological invariants that are protected by a combination of energy gaps and symmetries.  For example, topological insulators in two and three dimensions cannot be adiabatically connected to trivial insulators for as long as time reversal symmetry and charge conservation are satisfied.

The 2D topological insulator is characterized by a "helical" edge mode, which is an edge carrying two counter-propagating edge modes that are Kramers' partners of one another. When either TRS or charge conservation is broken, the edge is gapped. Interestingly, at an interface between these two types of gapping a localized Majorana mode is formed \cite{MajoranaQSHedge}. In 3D the surface carries a gapless Dirac cone. When TRS is broken the Dirac cone is gapped and the surface carries a Hall conductivity of $e^2/2h$. The surface may also be gapped by coupling to a super-conductor, and an interface between the two gapping mechanisms carries a chiral Majorana mode. Within the bulk a 3D topological insulator is characterized by the coupling of electric charge to magnetic monopoles. A magnetic monopole-antimonopole pair in the bulk may be created by drilling a thin solenoid into the bulk. This is the 3D analog of the insertion of flux into an annulus in a 2D setting. The void in which the solenoid is positioned is now surrounded by a gapless finite-size Dirac cone. When current flows in the solenoid, it creates a magnetic field pattern that corresponds to a monpopole-antimonopole pair. However, when the current is turned on, the time dependence of the magnetic field creates an electric field, which - due to the half integer Hall conductivity - makes a half-integer charge-dipole form on the two ends of the solenoid. This is a consequence of the Witten effect.

\subsection{What do we mean by "fractional"?}

It is "fractional" as in "the Fractional Quantum Hall effect", defined by the observation of the quantized Hall conductivity $\sigma_{xy}=\nu\frac{e^2}{h}$ with $\nu=\frac{p}{q}$ being a fraction. As mentioned earlier, it may be proven that with a single ground state on a torus $\nu$ must be an integer. Moreover, it may also be shown that for non-interacting electrons $\nu$ must be an integer \cite{TKNN,ThoulessNiu}. Thus, the fractional quantum Hall effect must originate from interactions, and must involve a degeneracy of the ground state on a torus \cite{Wen}. In fact, the degeneracy must  be a multiple of $q$. This is a remarkable observation. Spectral degeneracies accompany quantum mechanics from the early beginning, e.g., in the atomic shell structure. However, in usual cases they originate from a symmetry, e.g., a symmetry to rotation. In the present context, all symmetries (other than charge conservation) are broken, and yet the ground state of a torus in the thermodynamic limit is degenerate.

This degeneracy is closely related to the notion of a fractional charge. The relation between the two may be seen by constructing a torus out of two annuli glued together. Let us start with a single annulus in a FQH state $\nu=p/q$. The adiabatic insertion of a flux quantum $\Phi_0\equiv hc/e$ into the hole of the annulus keeps the system in an eigenstate (as guaranteed by the adiabatic theorem and the energy gap), yet transfers a charge $e\nu$ from one edge of the annulus to the other. The resulting eigenstate, then, is a state with a localized lump of charge of fractional value $p/q$. It may in fact be shown that the smallest value of the fractional charge $e^*$ is at most $1/q$. A bulk quasi-particle of fractional charge, the Laughlin quasi-particle, is obtained in the limiting case where the internal edge of the annulus is shrunk to small size.

The two annuli that we glue together to make a torus have altogether four edges, all of which we assume very large. All four carry  gapless chiral modes which may be excited by local perturbations. The spectrum of the edges is split to several {\it topological sectors}. In the simplest cases of abelian FQH states, the topological sectors are characterized by the value of the fractional part of the charge on the edge. Perturbations that do not couple the edges to one another and do not excite the bulk cannot change the topological sectors of the edges. Moreover, as long as the bulk is pristine and does not hold any quasi-particles, the topological sector of one edge determines the topological sector of the other edge of the same annulus, due to the constraint that the total charge on the annulus must be an integer. Each topological sector has a ground state and the energy difference between the ground states of different sectors is of the order of $1/L$, with $L$ the edge size.

To glue the two annuli we need to put them on top of one another, and couple their edges such that they are gapped in pairs - the exterior edges of the two annuli should gap one another, as should the interior edges. Since the magnetic field should be perpendicular to the surface at all points and the edge modes at the two annuli should be counter-propagating, the two annuli should be subjected to opposite magnetic fields. Alternatively, we could consider two annuli subjected to the same magnetic field, with one annulus populated by electrons and the other by holes. The latter realization does not require magnetic monopoles, and is thus experimentally accessible. A third realization is a 2D fractional quantum spin Hall state, where electrons of spin $\sigma=\pm$ form a fractional quantum Hall state of $\pm\nu$.  In all cases each edge carries counter-propagating gapless modes.

The gluing of the two annuli to one another is obtained by mutual gapping of the counter-propagating edge modes (for details see \cite{PhysRevB.91.245144}).
%The counter-propagating  modes at the exterior edges are mutually gapped, and so are the counter-propagating modes at the interior edges.
The simplest way to gap the edges is through back-scattering of electrons from one annulus to another. A back-scattering term that scatters electrons between the  counter-propagating modes on each edge makes the charge on each of these modes fluctuate, but does not affect the fractional part of that charge. When the fluctuations of the charge are large the dependence of the energy on the topological sector becomes exponentially small in $L$, leading to the toroidal ground state degeneracy. The degenerate ground states are then distinguished by the fractional part of the charge on each of the edges. Consequently, the number of degenerate ground states is an integer multiple of $1/e^*$. Ground state degeneracy on a torus and fractional quantum numbers are then tied hand in hand. In fact, tied to these two notions is another one, that of fractional statistics. When fractionally charged quasi-particles adiabatically encircle one another, they acquire a phase that is not necessarily a multiple of $2\pi$.

The counter-propagating edge modes may be gapped also by coupling to super-conductors, with non-abelian defects (parafermions) forming at the interface between different gapping mechanisms \cite{Barkeshliparafendleyons1, Barkeshliparafendleyons2,LindnerParafendleyons, ClarkeParafendleyons, Chengparafendleyons, Vaeziparafendleyons}.  These are reviewed elsewhere in this volume, in \cite{JasonPaul}.

\subsection{A short review of some technical aspects}

As the title of this subsection suggests, it will not be long. Details may be found in many books and papers, for example \cite{Wen,Levin2012}.

Bulk-edge correspondence suggests that some properties of topological systems may be analyzed either in terms of the bulk or in terms of the edge. The bulk description is most convenient for understanding the notion of  quasi-particles, their charge and statistics. The bulk of a  two dimensional abelian topological system is described in terms of a $N$-component $U(1)$
Chern-Simons theory whose Lagrangian density is of the form
\begin{equation}
L_{B} = \frac{{K}_{IJ}}{4\pi} \epsilon^{\lambda \mu \nu} a_{I \lambda} \partial_\mu a_{J \nu}
- \frac{1}{2\pi} \tau_I \epsilon^{\lambda \mu \nu} A_{\lambda} \partial_\mu a_{I \nu}
\label{kmat}
\end{equation}
where ${K}$ is a $N \times N$ symmetric, nondegenerate integer matrix and $\tau$ is a $N$ component
integer vector. The matrix ${K}$ is affectionately called the ``$K$-matrix'' and $\tau$ is known as the ``charge vector.'' The gauge fields $a_{I\lambda}$ are internal degrees of freedom while the vector potential $A_\lambda$ is externally applied.

In this formalism, the quasiparticle excitations in the insulator can be described
by coupling (\ref{kmat}) to bosonic particles which carry \emph{integer} gauge charge $l_I$
under each of the gauge fields $a_I$. Thus, each quasiparticle excitation corresponds to a
$N$ component integer vector $l$. The physical electric charge of each excitation is
given by
\begin{equation}
q_l = l^T {K}^{-1} \tau
\label{charge}
\end{equation}
while the mutual statistics associated with braiding one particle around another is given by
\begin{equation}
\theta_{ll'} = 2\pi l^T {K}^{-1} l'
\label{stat}
\end{equation}
The statistical phase associated with exchanging two identical particles is $\theta_l = \theta_{ll}/2$.
The quasiparticle excitations which are ``local''  -- that is, composed out of the
constituent electrons -- correspond to vectors $l$ of the form $l = \mathcal{K} \Lambda$
where $\Lambda$ is an integer $N$ component vector.

Examining (\ref{charge}), (\ref{stat}) we can see that the local excitation with $l = {K} \Lambda$
carries electric charge $\Lambda^T \tau$ and has exchange statistics $\theta = \pi \Lambda^T {K} \Lambda$.
This observation can be used to deduce a constraint on the possible values for ${K}, \tau$:
\begin{equation}
\tau_{I} \equiv K_{II} \text{ (mod $2$)}
\label{paritycond}
\end{equation}

 Much of our discussion below concerns the stability of edges, as well as the operations that create and annihilate electrons or quasi-particles on the edges. To that end, we note that for a two dimensional abelian system the action of the gapless edge modes is \cite{Wen}

\begin{equation}
S=\int dxdt \left [ \partial_t\phi^TK\partial_x\phi - \partial_x\phi^TV\partial_x\phi +\tau^T\partial_0\phi A_x\right ]
\label{edgeaction}
\end{equation}
Here $\phi$ is an $N$-dimensional vector of bosonic fields, and the $K$-matrix - the same one we encountered in the bulk - determines the commutation relations between the fields,
\begin{equation}
\left [\partial_x\phi_i(x),\phi_j(x')\right ] =2\pi \delta(x-x')\left (K^{-1}\right )_{ij}
\label{commrelation}
\end{equation}
The  charge vector $\tau$ determines the coupling of the fields $\phi$ to the electromagnetic vector potential $A$. We use a gauge where $A_0=0$.

Charged excitations are created on an edge by operators of the form $e^{i l^T\phi}$, where $l$ is an integer valued vector. The charge created by such an operator is again $l^TK^{-1} \tau$. Thus, an excitation created by $l=K\lambda$ with $\lambda$ an integer valued vector is again an excitation that creates an integer number $\lambda^T \tau$ of electrons.
%Excitations that are not of this form may carry fractional charges. The mutual statistics of two excitations $l_1,l_2$ is $2\pi l_1^T K^{-1}l_2$, and
The smallest charge of a quasi-particle, bulk or edge,  is $e^*\equiv \min_l \left (l^TK^{-1}\tau\right )$.

The number of distinct fractional quasi-particles may be counted by counting the number of linearly independent vectors $l$ that do not differ from one another by electronic vectors $K\lambda$. This number is the determinant of $K$.

Interactions affect edge modes in several ways. Interactions that do not involve counter-propagating edge modes affect the velocity matrix $V$, and hence the energy dispersion, but do not gap modes. Interactions that involve  counter-propagating edge modes may induce energy gaps. These interactions are expressed by terms of the form $\cos(\lambda^TK\phi)$ such that $\lambda^T t=0$ for charge conserving processes, or an even number, when a coupling to a super-conductor is involved. Each such term is able to gap a pair of counter-propagating edge modes. In order create an energy gap at the edge it is essential that $\lambda^T K\lambda=0$ such that the field $\lambda^T K\phi(x)$ commutes at different values of $x$ and may be pinned to a uniform value.

In a fractional topological insulator the edge may consist of several pairs of edge modes, in which case several cosine terms are needed, with several vector $\lambda^{(j)}$. These vectors have to satisfy  $\lambda^{(j)T}K\lambda^{(j')}=0$ for all $j,j'$.

\section{Fractional Topological Insulators in two dimensions \label{twod}}

We start with two simple examples for a fractionalized topological insulator in two dimensions,  continue to a general classification of abelian fractional topological insulators in 2D, analyze the stability of their gapless edge to perturbations that do not break time reversal symmetry and conclude with a few comments on non-abelian systems. We find it more convenient to carry out the entire discussion in terms of the edge description.

\subsection{Two simple examples of Fractional Topological Insulators in two dimensions}

Arguably the simplest example of a fractional topological insulator is that of the Fractional Quantum Spin Hall effect, obtained when electrons of spin-up form a quantum spin Hall state of filling factor $\nu$ and electrons of spin-down form a fractional QSHE state of filling $-\nu$ (see \cite{Levin:prl09, Bernevig:prl06}). For other examples of a fractionalized quantum spin Hall state see \cite{Young2008}). When put on a torus, the ground state degeneracy is the square of that obtained for each of the directions of the spin. When put on a plane, the edge carries two  counter-propagating sets of edge modes which transform to one another under time reversal. The edge modes may be gapped by a Zeeman field that scatters electrons between them. However, even when the edges are gapped the system is not topologically trivial as its bulk still hosts quasi-particles with fractional charge and statistics.
%The stability of the edge modes to perturbations that do not break time reversal symmetry is a more subtle issue that will be discussed in details below.

%At first sight, it is very difficult to realize this system experimentally, since it requires electrons of the two spin directions to be subjected to two opposite large magnetic fields, and to interact only through an equal-spin interaction. Further examination suggests simpler routes, however.

A system that may realize physics similar to that of the fractional QSHE is an electron-hole double layer where the electrons and the holes are of equal density, subjected to the same magnetic field, and form FQHE states of opposite signs\cite{LindnerParafendleyons, ClarkeParafendleyons}. Although the time reversal (TR) symmetry is replaced here by particle-hole symmetry, the physics, at least at low energies, should be similar. Alternatively, such a state may be created in a set of coupled quantum wires of varying spin-orbit coupling strength\cite{PhysRevB.91.245144, PhysRevB.90.115426}.

Ideally, the fractional QSHE would have no interaction between electrons of opposite spins, such that the ground state would be a straightforward product of wave functions for both spin direction. In real-life electronic systems, however, there would also be interaction between electrons of opposite spins.
%The stability of the fractional QSHE states to such interactions was studied numerically
When the interaction couples only electrons of equal spin,  the bulk is obviously gapped. The energy gap protects the state from an infinitesimal inter-spin interaction, but the gap is not guaranteed to withold such an interaction all the way to an SU(2) symmetric point. Numerical studies of several models led to the several conclusions \cite{PhysRevB.90.245401,PhysRevB.85.195113}. First, the fractional QSHE state is stable to significant amount of inter-spin interaction, possibly even all the way to the SU(2) symmetric point. Second, among the competing phases, a phase that spontaneously breaks time reversal symmetry and forms a FQHE state of $2\nu$ or $-2\nu$ is a strong contender. And third, single particle terms that mix spin directions without breaking time reversal symmetry may in fact increase the gap and stabilize the fractional topological insulator phase.

A second relatively simple example of a fractional topological insulator originates from the exactly solvable model of the "Toric Code" \cite{Kitaev2003}. Although the original model is a spin model, it may be generalized to a charge-conserving model, in which bosons reside on a lattice. The bosons' Hamiltonian includes local charging terms and hopping terms. These terms are tailored to be all mutually commuting, which requires uncommon correlated hopping terms. The resulting state is gapped and hosts two types of fractionalized excitations - charged quasi-particles, carrying a fraction of the boson charge, and flux quasi-particles that are neutral. While each of the two types has bosonic statistics with respect to its peers, there is a semionic mutual statistics between the two types of quasi-particles \cite{PhysRevB.66.205104, PhysRevB.67.115108, Levin2011, Ruegg:prl12}

Wave functions and model interactions for 2D fractional topological insulators were constructed in (\cite{PhysRevLett.107.126803,lu2012,simon2015fractional}. Experimental consequences of 2D fractional topologcal insulatorr have been explored at \cite{PhysRevLett.108.206804}. A double-semion fractional topological insulator state is found to emerge in a 2D system by a slave boson analysis at \cite{Maciejko:prb13}. Diagnostics of the fractional 2D topological insulator via a chiral anomaly was discussed in \cite{PhysRevB.89.075137}.

\subsection{General classification of abelian fractional topological insulators in 2D\label{genclass}}

%The topological classification of two dimensional gapped systems of non-interacting electrons in terms of Chern numbers is refined when time reversal symmetry is imposed. While the symmetry dictates the vanishing of the Chern number, it also introduces the $Z_2$ classification to trivial and topological insulators and protects helical edge modes at an interface between the two. Is there an analogous state of affairs in the fractionalized realm?
The topological classification of time reversal invariant charge conserving systems is bound to be enlarged by fractionalization, since two bulk states that have a different set of fractional charges cannot be connected to one another without a closure of the energy gap, with or without time reversal symmetry imposed. We now review the classification of such states under time reversal symmetry \cite{Levin2012,neupert2011b,Mesaros2013,PhysRevB.87.245120}. %Furthermore, we will also examine the conditions for the existence of gapped states that may be adiabatically connected to one another when time reversal symmetry is broken, but are topologically distinct as long as time reversal symmetry is maintained.

We consider states made of spin-$1/2$ electrons, where time reversal symmetry satisfies
\begin{eqnarray}
&&T\psi_\uparrow T^{-1}=\psi_\downarrow \nonumber \\
&&T\psi_\downarrow T^{-1}=-\psi_\uparrow
 \label{TRfermions}
 \end{eqnarray}  such that $T^2=-1$. For abelian electronic systems, the edge is described by a set of $2N$ bosonic fields $\bf \phi$, and electronic excitations are of the form $e^{i\lambda^TK\phi}$. Of the $2N$ fields, $2M$ may describe pairs of electrons, created due to interactions. Consequently, the charge vector has the form $(0_M,t',t,t)$ where $t'$ is an even valued $M$-component vector.  Under time reversal, Eqs. (\ref{TRfermions}) require the $2N$ fields to transform according to
\begin{equation}
\mathcal{T}^{-1} \Phi_I \mathcal{T} = T_{IJ} \Phi_J + \pi \mathcal{K}^{-1}_{IJ} \chi_J
\label{gentedge}
\end{equation}
where the matrix $T$ and vector $\chi$ are
\begin{equation}
T = \bpm -\bf{1}_M & 0 & 0 & 0 \\
	     0 & \bf{1}_{M} & 0 & 0 \\
	     0 & 0 & 0 & \bf{1}_{N-M} \\
	     0 & 0 & \bf{1}_{N-M} & 0 \epm  \ , \ \chi = \bpm x \\ 0 \\ 0 \\ t \epm
\label{gensolT}
\end{equation}
The vector $x$ has $M$ components which are either zero or one.
%rder to complete our description of
%time reversal we need to specify $\chi$, which we will call the ``time reversal vector''
%(in analogy with the charge vector $\tau$). This form of the transformation is determined by the requirement that $\mathcal{T}^2=-1$.

The most general abelian topological state that is time reversal symmetric has a $K$-matrix of the form,
\begin{equation}
\mathcal{K} = \begin{pmatrix}
       0 & A & B & B \\
		   A^T & 0 & C & -C \\
		   B^T & C^T & K & W \\
		   B^T & -C^T & W^T & -K
			\end{pmatrix} \ , \ \tau = \begin{pmatrix} 0 \\ t' \\ t \\ t \end{pmatrix}
\label{gensolK}
\end{equation}

Here, the matrix $A$ is of dimension $M \times M$, the matrices $K$ and $W = -W^T$ are of dimension $(N-M) \times (N-M)$
and the matrices $B,C$ are of dimension $M \times (N-M)$.
%Similarly, $t'$ is of dimension $M$ and
%$t$ is of dimension $N-M$. Finally, the vector $x$ is some $N-M$ dimensional vector of $1$'s and $0$'s.

This complicated-looking matrix is best digested bite-by-bite. For $M=W=0$ this $\mathcal K$ matrix is a system we already discussed, two spin-polarized FQH states that are time reversed partners of one another. For $M=N$ the matrix describes a toric-code type state formed by clusters of pairs of electrons, which are bosons. In between, the $\mathcal K$ matrix describes coupling of these two types of systems. Note, when the matrices $B,C$ are non-zero electronic operators, i.e., operators of the type $e^{i\lambda_e^T K\phi}$ with $\lambda_e^T\tau=1$ deconstruct the electron into components from the "fermionic" fields and from the "bosonic" fields. This deconstruction is similar in spirit to what is done in slave-particle theories for fractional topological insulators, reviewed in details in \cite{FieteMaciejko2015}.

This description is redundant. Two systems whose $\mathcal{K}, T$ matrices as well as the vectors $\tau,\chi$ are all related by an $SL(2N,Z)$ transformation are the same system described by two different choices of fields $\phi$.

\subsection{Topological classification vs. edge modes}

The question of adiabatic connection of two FTIs may be considered in two different settings that yield different answers. The first is the question of adiabatically varying the Hamiltonian of a uniform system  in a compact geometry such that it is turned from one FTI to another. Two states that are topologically distinct from one another may be connected this way only by a closure of the energy gap. The second question regards the occurrence of gapless edge modes when  the Hamiltonian is varied from one FTI to another {\it as a function of position}. As an example, consider a transition from an FTI in which electrons of spin up(down) form a $\nu=\pm 1/3$ FQH state to a vacuum state, where both spin directions have $\nu=0$. As a bulk transition, a closure of the gap is essential, since the former state has nine degenerate ground states on a torus, while the latter has a single one. However, a system where the region of $x<0$ is in the former state and the region $x>0$ is in the latter does not have to host a gapless edge mode along the line $x=0$. Indeed, the two counter-propagating edge modes that flow along the $x=0$ line in the absence of  inter-spin coupling may be coupled and gapped.

In contrast to a topological insulator of non-interacting electrons, an FTI does not have to carry a gapless edge mode in order to be topologically distinct from the vacuum. In the next subsection we examine the conditions under which such an edge mode exists.

\subsection{Stability of the gapless edge modes to TRS perturbations}

\subsubsection{The question and the answer}

 If  an interface of a fractional topological insulator with a vacuum does not have to host a gapless edge mode, when would such a mode be guaranteed to exist? And when would it be necessary to have time reversal symmetry and charge conservation in order to guarantee that the edge is gapless?

 For a topological insulator of non-interacting electrons edge modes are never guaranteed to exist - they can always be gapped by a Zeeman field. Once TR symmetry and charge conservation are imposed, edge modes exist for topological insulators and do not exist for trivial ones.
 Gnerally, topological states of matter that conserve charge and have non-zero charge/thermal Hall conductance are guaranteed to have gapless modes, to carry the charge and energy currents involved. For systems that have no Hall conductivity with no symmetry other than energy conservation (i.e., systems where charge may be exchanged with super-conductors), a gapped edge is possible if and only if there exists a subset of quasiparticle types $M$ such that (1) all the quasiparticles in $M$ have trivial mutual statistics (i.e., $m_i K^{-1}m_j=0$, with $m_{i,j}$ the integer-valued vectors that correspond to two quasi-particles that belong to $M$, and (2) every quasiparticle that is not in $M$ has nontrivial mutual statistics with at least one quasiparticle in $M$\cite{Levinedgenosymmetry}.

When charge conservation is imposed, a different condition emerges. Recall that the edge modes of an integer quantum spin Hall state with a spin Hall conductance of $\nu$ may be gapped without breaking time reversal symmetry for even $\nu$ and are protected by this symmetry for odd $\nu$ (thus reducing the topological classification of 2D TR symmetric insulators from $Z$ to $Z_2$). The gappability of edge modes of fractional TIs also depends on the parity of an integer number. Here the number is $\frac{1}{e^*}\tau^T \mathcal{K}^{-1} \chi$, where $e^*$ is the smallest charged excitation in the system
(in units of $e$). More specifically, the FTIs have protected edge modes if and only if
\begin{equation}
\exp\left [\frac{i\pi}{e^*}\tau^T \mathcal{K}^{-1} \chi\right]=-1
\label{paritytest}
\end{equation}
The number $\frac{1}{e^*}\tau^T \mathcal{K}^{-1} \chi$ is an integer, and the condition (\ref{paritytest}) amounts to it being odd \cite{Levin2012}.

Before reviewing the way this expression is derived \cite{Levin2012} we look at simple cases. The simplest case is probably that of two time-reversed fractional quantum Hall states ($M=0$ and $W=0$ in Eq. (\ref{gensolK})). In that case the criterion for protected edge modes reduces to $\nu/e^*$ being odd. For fractional quantum Hall states of the Jain series, the parity of this number is the parity of the number of edge modes for the FQHE state of $\nu$. However, this is not generally true: for the abelian strongly paired $\nu=1/2$ state there is only a single edge mode, yet the ratio $\nu/e^*=2$ and the state may be gapped with no breaking of time reversal symmetry.
When the system consists of purely bosonic insulators $M=N$, the existence of protected edge modes depends on the vector $x$, i.e., on the precise way the bosons transform under time reversal.

\subsubsection{Pair switching, or Kramers' pump, for non-interacting topological insulators}

The physics behind the condition (\ref{paritytest}) combines together an essential aspect of topological insulators - Kramers theorem - with an essential aspect of fractionalized phases - ground state degeneracies and topological sectors. Let us expand shortly on these ingredients.

Gapless edge modes in non-interacting topological insulators are a consequence of two ingredients - the Kramers degeneracy of eigenstates of time-reversal-symmetric systems when the total spin is half-integer and the so-called $Z_2$ pumping of local Kramers' degeneracy between the edges of the annulus, associated with the threading of half a flux quantum through the annulus' hole \cite{KaneZ2}. Kramers' theorem states, first, that for a TRS system, applying a global transformation of time reversal $T$ to an eigenstate $|\psi\rangle$ gives an  eigenstate $T|\psi\rangle$ with the same energy; and second, that for half-integer spin $\langle\psi|T\psi\rangle=0$  while for an integer spin the two states do not need to be orthogonal.

\begin{figure}
\begin{center}
\includegraphics[width=\linewidth]{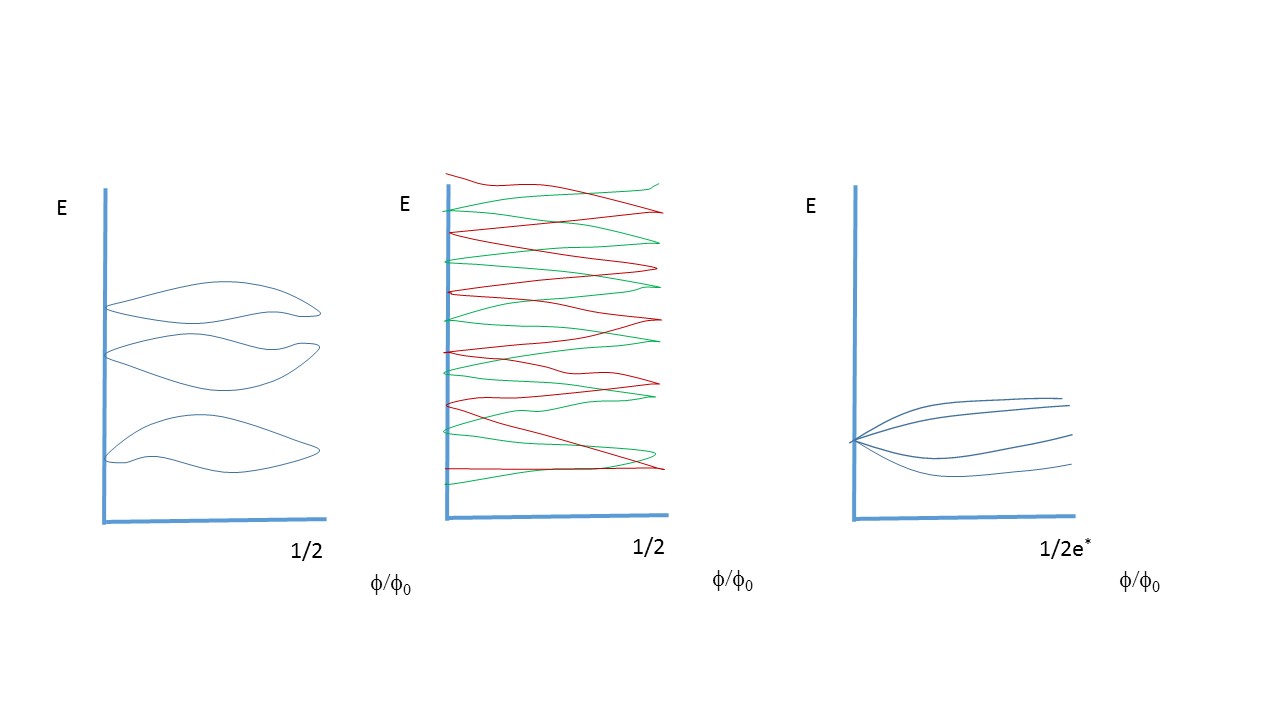}
\caption{ \label{nopairswitch}%
(a) Single particle spectrum as a function of flux for a finite size time-reversal-symmetric system. States are grouped to Kramers pairs at zero and half flux quantum. When the spectrum is bound and other spectral degeneracies do not exist, Kramers partners are not switched between the two time reversal symmetric points.
(b) Single particle spectrum of edge modes of topological insulators. When the distance between the edges is very large, spectarl degeneracies of states at different edges are not avoided, and therefore a zig-zag pattern may exist on both edges (denoted by different colors) in the energy gap between two bulk bands.
(c) The many body spectrum when interactions are important (e.g. a fractional topological insulator). Both edges have a Kramers degeneracy at zero flux, and do not have it at a flux of $1/2e^*$. Only states of the same topological sectors are drawn.
}
\end{center}
\end{figure}

A TR-symmetric system placed on an annulus that is threaded by magnetic flux $\phi$ has special values of the flux $\phi=\frac{n\Phi_0}{2}$ where the flux does not break TR symmetry and Kramers' degeneracy occurs. Now, let us examine the single-electron spectrum as a function of the flux threading the hole of the annulus. We first make two assumptions - that no accidental degeneracies occur when $\phi\ne0,\frac{\Phi_0}{2}$, and that the spectrum is bounded. With these two assumptions, the flux dependence of the eigenenergies looks as in Fig. (\ref{nopairswitch}(a)), i.e., the pairs of degenerate eigenstates cannot be switched between $\phi=0,\frac{\Phi_0}{2}$. Two states that are Kramers' partners at $\pi=0$ evolve to be Kramers' partners at $\phi=\frac{\Phi_0}{2}$.  For a switching to occur, a degeneracy (an "unavoided" crossing, to use a somewhat clumsy expression) must occur somewhere between $\phi=0,\frac{\Phi_0}{2}$. For that to happen, two seemingly contradicting conditions must be satisfied. First, the flux should be able to affect the energy of the switching state by an amount comparable to the level spacing, requiring the states not to be localized. And second, energy crossings at arbitrary values of the flux should be allowed to stay unavoided, thus the possibility of states that are spatially separated from one another should be allowed. The resolution for this seeming contradiction comes in the form of edge modes, in which states extend around the entire perimeter and are thus affected by the magnetic flux, but decay into the bulk and thus may have unavoided crossings with states of the other edge. The decay into the bulk requires a bulk gap, and thus pair switching occurs at energies where the bulk is gapped, i.e., between bands. For each of the edge modes, a "pair-switching" occurs, in which the energy vs. flux plot forms a zig-zag pattern (see Fig. (\ref{nopairswitch}(b)). There can be either none or one edge mode per edge, since a pair of zig-zags on the same edge would not be protected against the splitting of the degeneracies.
%This is the manifestation of the $Z_2$ classification of the non-interacting topological insulators.

When edge modes exist together with a gapped bulk in a TR-symmetric state, it is possible to define a "local time reversal", that is, a transformation that reverses the time in one edge of the annulus (We formulated such a transformation in subsection (\ref{genclass})). When that edge has a half-integer spin, the transformed state would be degenerate and orthogonal to the initial state, and the edge carries a "local Kramers' degeneracy". Each edge may then be characterized by a binary $Z_2$ index indicating whether or not it carries a local Kramers' degeneracy. In the case where the edge spectrum exhibits pair switching, an insertion of half a flux quantum into the hole of the annulus (a change of $n$ by one), switches that index on both edges. This may be understood by considering (say, for $n=0$) an edge where all Kramers pairs below a certain energy are occupied, and all pairs above that energy are empty. Now consider the pair with the highest energy. As the flux is varied and $n$ changes by one, one of the electrons of that pair becomes the sole occupant of a Kramers pair that has been previously empty. That electron has two possible degenerate states to be in, hence a local Kramers' degeneracy.
This consideration leads also to the $Z_2$ topological robustness of the 2D topological insulator. Since the entire system cannot exhibit pair switching, when one edge does, so does the other one. Then, no TR-symmetric local perturbation that acts on a single edge can change that.
%
%Perhaps the defining feature of 2D topological insulators is their $Z_2$ pumping. For non-interacting electrons the $Z_2$ pumping is a consequence of the "pair-switching" associated with the insertion of half a flux quantum $\Phi_0/2$ through the annulus' hole. When the flux through the hole is an integer number of $\Phi_0/2$ the flux does not break time reversal symmetry, and the spectrum is then composed of Kramers pairs of degenerate eigenvalues. Assuming the absence of any other accidental degeneracies, {\it all} states in the spectrum must conform to one of two statements: either they keep the same Kramers' partner at $\phi=0$ and $\phi=\Phi_0/2$ (this case corresponds to a trivial insulator) or they switch their partners between these two values of the flux (this case corresponds to topological insulators). Since the effect of flux in the hole on the energy of an eigenstate is of the order of $1/L$, the latter behavior of "pair switching" at the Fermi energy is not compatible with

\subsubsection{Pumping Kramers' degeneracy in fractional topological insulators}

Two of the considerations explained above are going to be modified when we deal with fractional topological insulators - the relevant flux unit will be $hc/e^*$, and the zig-zag pattern will not occur. We will replace the zig-zag pattern by a different, though related, consideration, resulting in a weaker statement.

Fractionalized phases, e.g., the fractional quantum Hall state, have ground state degeneracies when put on a torus in the thermodynamical limit, with the finite-size splitting of the degeneracies being exponentially small in the size of the torus. To create a single annulus out of a torus, we may perform a cut along one of the torus' two circumferences. For the annulus the ground states are not degenerate anymore. When the annulus carries a gapless mode, the energy separation between the (generically single) ground state and the low energy excited states scales like $1/L$ with $L$ the annulus circumference. Low energy edge excitations are grouped into {\it topological sectors}, where two excited states belong to the same sector if they are related by a local operator, i.e., an operator that involves electrons within a confined part of space.
 %For the fractional quantum Hall effect, two states belong to the same sector if the fractional part of the charge on the edge is the same for both.
  The topological sector of an edge may be identified by the Berry phases accumulated by quasi-particles that encircle it. It may not be changed by local excitations. Rather, it is changed by a transfer of a fractionalized excitation from one edge to another.
The number of topological sectors of a fractionalized phase on an annulus is the same as the number of degenerate ground state it has on a torus.\cite{Wen}

Topological sectors affect the dependence of the energy on the magnetic flux in an annulus geometry. The spectrum as a whole is periodic with a period of one flux quantum. Furthermore, for TR-symmetric states, the spectrum is even with respect to the flux. Both statements are not necessarily true, however, when considered for each topological sector. In order to focus on the effect of TR-symmetry on the spectrum we need to modify the arguments we used above in such a way that they explore spectral degeneracies of states from the {\it same} topological sectors. To that end, it is important to note that while the insertion of a flux quantum $\Phi_0$ generally transforms the annulus edges from one topological sector to another, the insertion of  $1/e^*$ flux quanta necessarily brings the two edges back to the topological sector they started from. The insertion of such a flux corresponds to the operator
\begin{equation}
U_*\equiv \exp{\frac{i}{e^*}{\vec t}\cdot{\vec\phi}}.
\label{ustar}
\end{equation} This is a local operator, made of electronic operators confined to the region of the edge. This is so since we the vector
\begin{equation}
\frac{1}{e^*}K^{-1}{\vec t}
\label{integers}
\end{equation}
is an integer valued vector.

Thus, the comparison we carried out in the non-interacting case between local Kramers degeneracies at flux values of $0$ and $\Phi_0/2$ has to be carried out now between flux values of $0$ and $\Phi_0/2e^*$. In the fractional case, which is inherently a many body problem,  we cannot address the zig-zag patterns of single particle states. Thus, we examine the pumping of local Kramers degeneracies. We assume that the number of electrons in the system is even, such that Kramers' theorem does not enforce Kramers degeneracy. We further assume that at zero flux there is no ground state degeneracy and $|\psi_0\rangle={\cal T}|\psi_0\rangle$, and ask what are the conditions under which there will be such a degeneracy at a flux $\Phi/2e^*$. Since the change of energy values that result from  the insertion of $1/2e^*$ flux quanta is of order $1/L$, pumping of Kramers degeneracy is an indication that time reversal symmetry adds low energy states to the many body spectrum. Although it is not a proof of the existence of a gapless edge, it suggests that such an edge exists. As we see in the next section, in three dimensions this is not the only possibility.

To establish Kramers' degeneracy, we need to find an anti-unitary operator $\cal A$ such that ${\cal A}^2=-1$ and ${\cal A}$ commutes with the Hamiltonian. Under these conditions, if $|\psi\rangle$ is an eigenstate of the Hamiltonian with an energy $E$, so is ${\cal A}|{\psi}\rangle$, and the two states are orthogonal. The time-reversal operator ${\cal T}$ cannot serve as the ${\cal A}$ here - for an even number of electrons it satisfies ${\cal T}^2=1$. Since the idea is to have a "local Kramers degeneracy" we need to multiply $\cal T$ by a unitary operator that "time reverse" a single edge. When the flux inserted is $1/2e^*$ flux quanta, a single edge would be "time reversed" by an operator that corresponds to the insertion of $-1/e^*$ flux quanta, i.e., by $U_*$. So, can we identify ${\cal A}={\cal T}U_*$? Let us see: $({\cal T}U_*)^2={\cal T}U_*{\cal T}U_*={\cal T}^{-1}U_*{\cal T}U_*$. But ${\cal T}^{-1}U_*{\cal T}$ is nothing but the time-reversed $U_*$, and by the recipe we defined above, ${\cal T}^{-1}U_*{\cal T}=U_*^{-1}\exp{\frac{i\pi}{e^*}t K^{-1}\chi}$. Consequently, we find a condition for the formation of local Kramers' degeneracy on both edges when the flux is $1/2e^*$. It is Eq. (\ref{paritytest}). When it holds, the edge cannot be gapped without TR symmetry being broken.

When $\frac{1}{e^*}\tau^T \mathcal{K}^{-1} \chi$ is even TR symmetry does not  protect the edge from being gapped. Furthermore, an explicit set of TR symmetric perturbations that gaps the edge may be constructed \cite{Levin2012}.

\subsection{Weak spontaneous breaking of time reversal symmetry}

Interesting phenomena occur at the end points of regions where counter-propagating edges are gapped by mutual coupling. A prominent example is the formation of non-abelian defects - parafermions - at interfaces between different gapping mechanisms, e.g. gapping by back-scattering and gapping by a super-conductor \cite{Barkeshliparafendleyons1, Barkeshliparafendleyons2,LindnerParafendleyons, ClarkeParafendleyons, Chengparafendleyons, Vaeziparafendleyons}. These parafermions are reviewed elsewhere in this volume \cite{JasonPaul}, and we therefore do not review them here.

Another example is the spontaneous breaking of time reversal symmetry at the end point between a gapped region and a region where the edges are gapless, known also as "weak symmetry breaking" \cite{PhysRevB.88.245136,Mross2015}.
The criterion (\ref{paritytest}) does not lend itself to the $Z_2$ structure that characterises non-fractionalized systems, in which $e^*=1$. In particular, consider two spin-conserving quantum spin Hall states, with the first being the non-interacting $\nu=\pm 1$, and the second being the interacting $\nu=\pm 1/2$. The former has a K-matrix of $\sigma_z$, with a charge vector of $\tau=(1,1)$ and time reversal vector of $\chi=(1,0)$. The second has a K-matrix of $8\sigma_z$, a charge vector of $(2,2)$ and a zero time reversal vector. As a consequence, the edge of the first is stable to perturbations that keep TRS, while the latter may be gapped without breaking TRS. Furthermore, an edge made of a combination of the two is gappable as well, as is an edge made of two copies of the $\sigma_{xy}=\pm1/2$ insulator.

\begin{figure}
\begin{center}
\includegraphics[width=\linewidth]{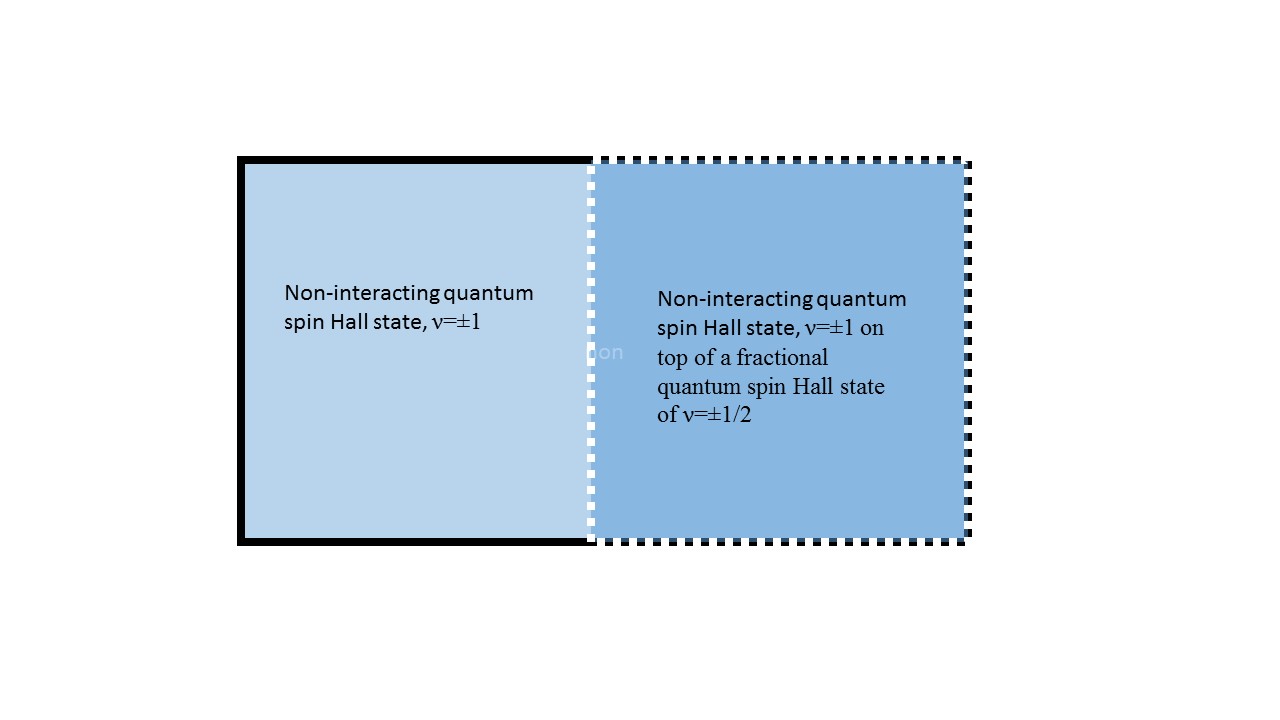}
\caption{ \label{weakbreaking}%
Weak spontaneous breaking of time reversal symmetry. The entire rectangle is a non-interacting quantum spin Hall state. The right half is covered by a $\nu=\pm 1/2$ fractional topological insulator. The black sides carry a gapless helical state of non-interacting electrons. In the dashed black-white sides the edges of both topological insulators are gapped together, with no breaking of TRS. In the dashed white line the edge of the fractional state is gapped with no breaking of TRS. At the intersections of all three lines local magnetic moments must form.
}
\end{center}
\end{figure}

Now consider a rectangular sample of a non-interacting 2D topological insulator, with half of the rectangle covered by a $\nu=\pm 1/2$ fractional topological insulator (see Fig. (\ref{weakbreaking})). By the criterion (\ref{paritytest}) it is possible to introduce back-scattering terms that conserve charge and respect TRS and leave gapless only the left half of the rectangle's circumference. When that happes, an electron that moves counter-clockwise along the gapless edge has to be scattered back when it reaches the gapped region. However, such a scattering requires a spin-flip, which in turn requires a local spontaneous breaking of TRS. At the interface between the two TRS mechanisms of gapping - the one that gaps the $\pm 1/2$ region alone and the one that gaps both systems together - a local magnetic moment must develop.

\subsection{Non-abelian fractional topological insulators in two dimensions}

The simplest example of a non-abelian fractional topological insulator is a fractional QSHE made of the non-abelian states that are time reversed partners of one another. The edges of such states carry counter-propagating edge modes of fractional central charge. The edges may be gapped by a backscattering that break TRS. The stability to perturbations that do not break TRS is a more subtle issue, which at present is not conclusively understood (see however \cite{Cappelli2015}).

A different method \cite{Koch:prb13} for constructing non-abelian models relies on the decomposition evident in Eq. (\ref{gensolK}) of the system to a fermionic part coupled to a bosonic part. If the electrons are decomposed into a charged fermionic part and a neutral bosonic part, the bosons are made to construct a non-abelian phase and the fermions to occupy a topological insulator band structure, the resulting phase is a non-abelian topological insulator. This method essentially replaces the upper-left quadrant of $\cal K$ in Eq. (\ref{gensolK}) by a non-abelian system, and the lower-right quadrant by $\sigma_z$.

\section{Two dimensional fractional topological insulators on surfaces of three dimensional topological insulators \label{twodonthreed}}

 The zig-zag pattern of single particle energies as a function of magnetic flux, introduced in the previous section (Fig. (\ref{nopairswitch}(b))), was sufficient to conclude that protected gapless edge modes may exist on the edges of insulators of non-interacting electrons. Unfortunately, when interaction was turned on, we lost the zig-zag pattern, and had to replace it with the $Z_2$ pumping argument (Fig. (\ref{nopairswitch}(c))), whose conclusion was  that at the bottom of the spectrum there must be at least four states (two on each edge) whose mutual energy separation is bounded by an energy that scales like   $1/L$.

 For non-interacting electrons in 3D this is explicitly seen through the dependence of the Dirac cone spectrum on the two fluxes that may thread the torus. Assuming the torus' two circumferences to be equal, for simplicity, the spectrum is
 \begin{equation}
 E(k_x,k_y)=\frac{2\pi v}{L}\sqrt{(n_x-\frac{\Phi_x}{\Phi_0})^2+(n_y-\frac{\Phi_y}{\Phi_0})^2}
 \label{diracspectrum}
 \end{equation}
with $n_x,n_y$ being integers. The degeneracies at half a flux quantum are evident.

  In the presence of surface interactions a new possibility emerges, anticipated in (\cite{Levin2011}) and realized in \cite{BondersonTO,WangTO,ChenTO,MetlitskiTO}, by which the low lying states are degenerate ground states that result from topological order on the surface. In this scenario the surface forms a {\it gapped topologically ordered state that is charge conserving and symmetric to time reversal}. Consider the "thickened torus" geometry, namely a system that is finite in the $z$-direction and has periodic boundary conditions in the $x,y$ directions. Such a system has two surfaces, and may be viewed as quasi-two dimensional. Since the word "quasi" is only quasi-well defined, let us say explicitly what we mean - with both surfaces included and gapped, the system is a two dimensional fractional topological insulator.

\begin{figure}
\begin{center}
\includegraphics[width=\linewidth]{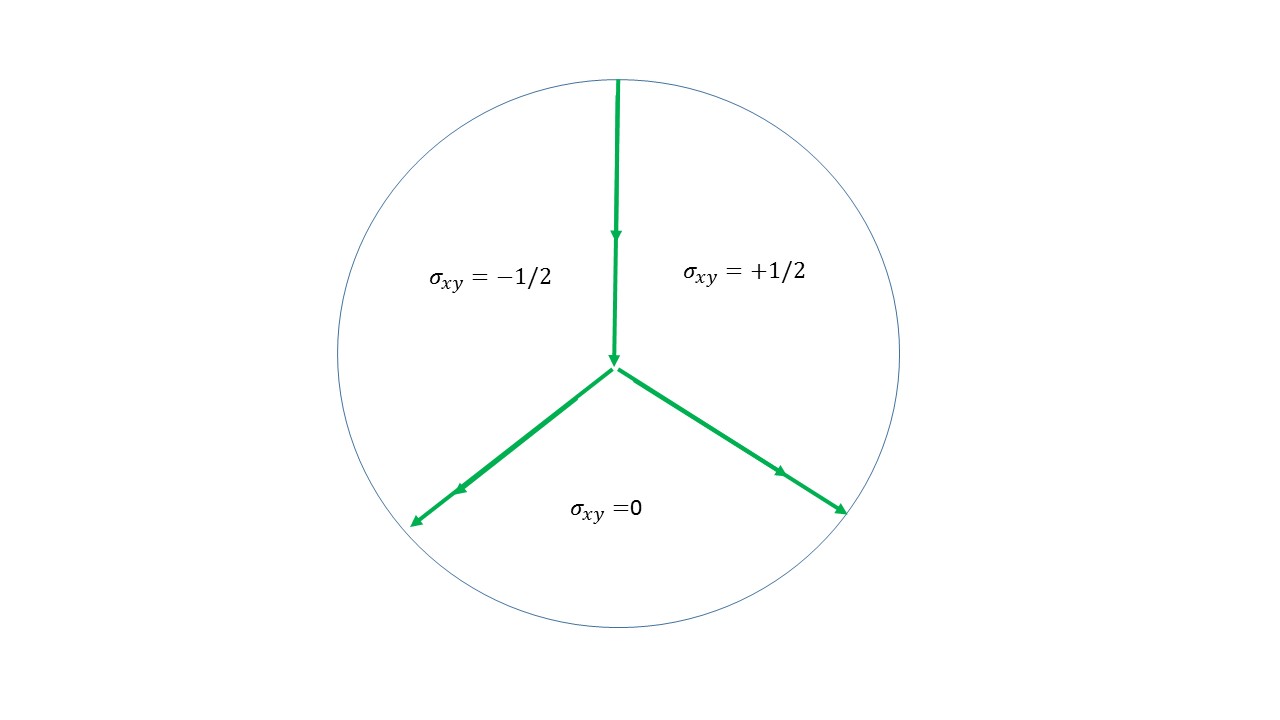}
\caption{ \label{peacesymbol}%
The symmetrically gapped surface of a 3D topological insulator must be non-abelian. To see that, look at the geometry above, where a surface of a 3D topological insulator is gapped in three different ways. Two regions are gapped by a ferromagnet of two opposite magnetization. They have Hall conductivities of $\sigma_{xy}=\pm 1/2$. The third is gapped without breaking time reversal symmetry. By symmetry, the vertical edge mode that carries $\sigma_{xy}=1$ and a central charge of one must split equally between the other two edge modes, thus yielding a fractional central charge, which implies a non-abelian state.
}
\end{center}
\end{figure}

  This FTI must be non-abelian, as can be seen from the following consideration (see Fig. (\ref{peacesymbol})): imagine a T-junction on the surface of a strong TI, where the junction divides the surface to three parts - one magnetized to carry a $\sigma_{xy}=1/2$ quantum Hall conductivity (in units of $e^2/h$); one magnetized to carry a $\sigma_{xy}=-1/2$ quantum Hall conductivity; and the third gapped in a way that is symmetric to TR. All arms of the junction must carry gapless modes. In between the two magnetized regions there is a single chiral fermion mode, carrying a Hall conductivity of one and a central charge of one. By symmetry, the interfaces between the magnetized regions and the symmetrically gapped regions must then carry a Hall conductivity and a central charge of one-half each. An edge with a fractional central charge implies non-abelian statistics.

 While gapping the two surfaces forms a fractional topological insulator, a single surface state may not be formed as a stand alone two dimensional gapped system. To "peel it off" from the three dimensional topological insulator, either time reversal symmetry of charge conservation have to be broken. Were that not the case, one would be able to peel off the external surface of the thickened torus, magnetize the internal surface to carry a half-integer Hall conductivity and be left with a two-dimensional gapped system of non-interacting electrons and fractional Hall conductivity, which is impossible.

\begin{figure}
\begin{center}
\includegraphics[width=\linewidth]{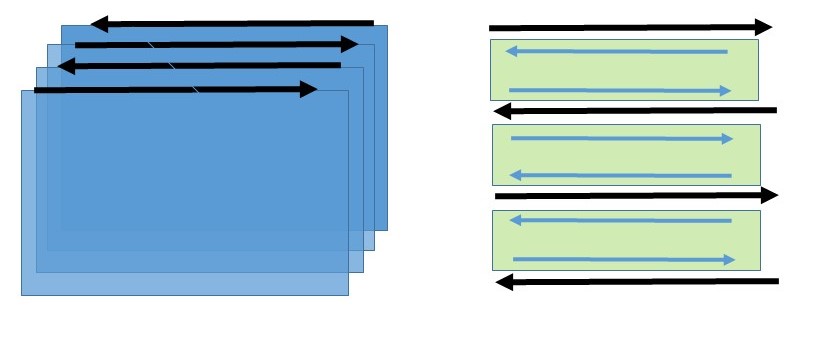}
\caption{ \label{Mrosssto}%
(a) The anti-ferromagnetic topological insulator \cite{AFTIMong} is composed of layers of integer quantum Hall states of alternating filling factors $\pm 1$. The edge states flow in alternating directions and their coupling leads to a single surface Dirac cone. (b) Top view on the surface. To gap the Dirac cone without breaking the symmetry, plates of fractional quantum Hall states of filling factors $\nu=\pm 1/2$ are positioned parallel to the surface. Each interface has two co-propagating edge modes of $\nu=1/2$ and a counter-propagating $\nu=1$ mode. Coupling the three gaps two modes and leaves behind a single neutral mode.
}
\end{center}
\end{figure}

 Several ideas were proposed to form such a gapped symmetric surface of a three dimensional topological insulator\cite{BondersonTO,WangTO,ChenTO,MetlitskiTO, STO}. As is commonly the case, much intuition may be obtained by considering the two dimensional surface as composed of many parallel and coupled one dimensional systems \cite{KaneWires,TeoKaneChains,STO}. While the surface of a strong topological insulator cannot be described this way, a close relative can - the anti-ferromagnetic topological insulator, which consists of a stack of two dimensional layers at alternating quantum Hall states of $\nu=\pm 1$ \cite{AFTIMong} (see Fig. (\ref{Mrosssto})(a)). This stack is not symmetric to time-reversal, but is symmetric to the anti-unitary combination $\tilde T$ of the anti-unitary time reversal and the unitary translation of half a unit cell. Since ${\tilde T}^2=-1$ it plays the same role that time reversal symmetry plays for topological insulators. The surface is then an array of chiral fermionic modes of $\nu=\pm 1$ QHE, i.e., of alternating directions. As long as the symmetry $\tilde T$ is not broken, the surface is gapless and its spectrum hosts a single Dirac cone.

 The advantage of this description is that the gapping of the surface is now translated to the problem of gapping an array of one dimensional chiral modes. Since each mode carries a Hall conductivity of $e^2/h$, its charged part may be gapped when it is coupled to two edge modes that originate from two "plates" of $\nu=\pm 1/2$. These two modes flow both in an opposite direction to the $\nu=\pm 1$ mode that they gap, and are located on its two sides (see Fig. (\ref{Mrosssto})(b)).  The Hall conductivity of the combined system is zero, but since two co-propagating $\nu=1/2$ edges attempt to gap a single $\nu=1$ edge, a neutral mode must remain. This scheme then turns the alternating array of $\nu=\pm 1$ edge modes into an alternating array of neutral modes, and the charged fermionic Dirac cone turns into a neutral one. The stripping of the Dirac cone off its charge now allows it to be gapped without breaking charge conservation, by Cooper pairing. The resulting state is a super-fluid of the neutral fermions Cooper pairs, with the pairing having a p-wave symmetry. As such, it has Majorana edge modes of the expected central charge $1/2$, and carry localized Majorana modes at the cores of vortices.

 It turns out that there are several procedures that may gap the surface of a strong topological insulator, with the resulting surface states being the "T-Pfaffian" and the "pfaffian-antisemion". The surfaces of bosonic topological insulators \cite{AshvinSenthil}, weak topological insulators \cite{Mross2015} and topological crystalline insulators \cite{QiFu2015} may be gapped as well to form states of topological order, but the resulting states are not necessarily non-abelian.

 %Similar conclusions hold for bosonic topological insulators \cite{AshvinSenthil}, topological superconductors \cite{MetlitskiTO2}, and topological crystalline insulators \cite{QiFu2015}.  (Not all topological systems, however, admit a symmetric gapped boundary \cite{WangSenthil2014}.)

 To conclude this section, we emphasize that the topologically ordered state here is confined to a single surface. In the thickened torus geometry, each of the surfaces must overall be in a vacuum topological sector.

 \section{Three dimensional fractional topological insulators\label{threed}}

 Thanks in a large part to the fractional quantum Hall effect, we usually associate fractionalized phases with two dimensions. Two dimensional systems are a natural playground for fractional statistics, since they provide a classification of trajectories of point particles in terms of winding numbers. However, the expansion of the world of topological systems from two to three dimensions, which was initiated by the discovery of the 3D topological insulators, naturally suggests a search for 3D fractional topological insulators\cite{Maciejko:prl10,Levin2011,Swingle:prb11,Maciejko:prb12,Maciejko:prl14,PhysRevX.4.041049,park2010dirac,Cho2012dyon}. Defining characteristics of such a phase may be expected to be a degeneracy of the ground state on the boundary-less three dimensional torus, a surface quantum Hall effect of $\nu=(pe^2)/2qh$ (with $p/q$ being a fraction), fractionally charged excitations in the bulk, mutual statistics between bulk point and loop particles and a coupling of fractional charges to monopoles in the bulk.

 A model for a 3D fractional topological insulator is most easily built on the foundations of a related problem whose solution we are familiar with, such as a 3D non-interacting topological insulator, a 2D fractional topological insulator, or a set of coupled one dimensional wires. Let us expand shortly on each of the these.

 %Arguably the simplest route towards such systems is a route that makes use of what we know about topological insulators of non-interacting electrons.
 In the same way that the fractional quantum Hall effect may be understood as an integer quantum Hall effect of composite fermions, which amount to electrons dressed by an even number of flux quanta, one may search for a description of a fractionalized 3D phase by decomposing the electron to a combination of a fermion and a boson, in which the fermion occupies a topological band, and the boson is gapped by interactions. For the system to be fractionalized, the bosonic subsystem should be characterized by topological order including excitations with fractional charge and statistics as well as ground state degeneracy on a three dimensional torus.

 An explicit exactly solvable model for such a system is constructed in \cite{Levin2011}. In the first step bosons are composed of tightly bound pairs of electrons, and are subjected to a charge conserving version of the toric code Hamiltonian. The fractionally charged excitation of charge $e^*$ that appears in the 2D model carries over to 3D, but the flux particle changes from being a point particle in 2D to a loop in 3D. Again, each of these two types of quasi-particles has trivial statistics with respect to its peers, but there is a non-trivial mutual statistics - when the charged point-particle "knots" the loop the wave function acquires a quantized phase. Other than this mutual statistics, the quasi-particles of the bosonic system do not interact with one another.

 The second step forms composite fermions of fractional charge $q=1+ke^*$ by attaching $k$ fractionally charged excitations of the bosonic systems to each electron. The composite fermions have non-trivial statistics with the flux quasi-particles, but fermionic statistics with one another. They can therefore be made to form energy bands of non-trivial topology, resulting in a topological insulator of fractionally charged fermions, with flux quasi-particles as high energy excitations.

 The combined system inherits the ground state degeneracy on the three dimensional torus from the bosonic subsystem. Under breaking of TR symmetry, the surfaces form a quantum Hall state of fractional Hall conductivity $q^2/2h$, with edges that have $q/e^*$ topological sectors. As a consequence, the attachment of monopoles to charges is different from that of the non-interacting case. A monopole-antimonopole pair of strength $1/e^*$ carries a charge of $q/2$.

 The model may be formulated for 2D and 3D. Interestingly, the set of parameters for which gapless edge modes may be shown to be guaranteed to exist is the same for both dimensions.

 Although the coupling to a super-conductor is harder to treat analytically, it is possible to see that its outcome is a gapping of the surface. Far less obvious is the form of an interface between a surface region that is gapped by a super-condtcor to one that is gapped by a magnetic field. In a way similar to that at which such an interface "halves" the edge mode of an integer quantum Hall state into a Majorana mode, one would expect a fractionalization of the edge mode of the fractional quantum Hall state of $q^2/h$. The precise way at which that happens is not simple to see, but it is presumably related to the notion of chiral parafermionic edge modes.

A bosonic 3D fractional topological insulator may be created by viewing each boson as a tightly bound singlet state of two fermions \cite{Maciejko:prl10,Swingle:prb11,Maciejko:prb12,Maciejko:prl14}. If now the fermions are subjected to periodic potential that gives rise to a strong topological insulator, one could search for a state in which the fermions occupy only the lower energy bands, yet satisfy the constraint of having either zero fermions (=zero bosons) or two fermions (=a singlet, i.e., a single boson) on each lattice site. Such a state has the fermions as high energy fractionalized excitations. The procedure is implemented in terms of a mean field $Z_2$ gauge theory. Similarly, electronic states may be created in terms of parton constructions, in which the electrons are decomposed into several partons, whose gluing is carried out by means of a gauge field.

 Another route to 3D fractional topological insulators starts from 2D layers which are stacked on one another \cite{LinLevin15,JianQi}. As long as the layers are uncoupled, the resulting ground state degeneracy on a 3D torus would scale exponentially with the number of layers. To reduce that degeneracy to a number that is independent of the number of layers, as we expect from a fractionalized 3D system, we must reduce the set of quasi-particles. This is done by forming bosons out of quasi-particles that reside in different layers, and condensing these bosons. Such a condensation requires a closure of the energy gap - it is a topological phase transition \cite{PhysRevB.79.045316}. But it is a "designed" phase transition, at which we have a good handle on the resulting state. By their condensation the bosons confine all the quasi-particles with which they have non-trivial statistics. The condensed bosons should be composites of quasi-particles of neighboring layers, such that they are - on one hand - local, and - on the other hand - building a three dimensional structure. The resulting three dimensional state may be constructed to have a small number of fractionalized quasi-particle types that neither condense nor get confined, thus forming a 3D fractional topological insulator.

 In the latter models non-trivial mutual statistics occurs also between vortex loops. This type of statistics is subtle. The phase accumulated when one loop winds around another by a trajectory that cannot be shrunk to a point without the loops touching one another crucially depends on the possible presence of a third loop that links the first two. The mutual loop statistics is fully defined only in terms of a three loop process.

 A different route to a 3D fractional topological insulator starts from a set of coupled wires \cite{Neupert2014,Sagi2014,PhysRevB.90.115426,PhysRevB.91.245144,PhysRevB.91.205141}. A 2D topological insulator may be constructed out of a set of parallel wires with an alternating sign of charge (electrons/holes) and with spin-orbit and tunnel couplings. The couplings may be tuned to bring the 2D system either into a trivial insulator phase, or into a topological insulator phase. They can also be tuned to have the system at the transition points between the two phases, in which the spectrum is gapless and consists of two Dirac cones.

 To construct a 3D topological insulator, layers at the critical state are stacked onto one another, and are coupled in such a way that Dirac cones gap in pairs, where each pair is composed of one Dirac cone of layer $i$ and another of layer $i+1$. The first and the last layer are then left gapless, with a single Dirac cone each - a 3D topological insulator. A similar procedure may be used to create a 2D system hosting Fibonacci anyons\cite{Mong,JasonPaul}.

 To make the topological insulator fractional, an interaction has to be added. The formulation of the quantum Hall effect in terms of coupled quantum wires \cite{KaneWires,TeoKaneChains} essentially gives a recipe for transforming an "integer" topological state into a "fractional" one. The recipe involves replacing the single electron tunneling from a wire to its neighbor by a process in which the electron tunneling is accompanied by creation of particular electron-hole excitations in the wires involved. When the recipe is applied here, it creates a fractional topological insulator.

The wires route is particularly convenient for analyzing the gapping of the surface by coupling to a super-conductor, finding that the interface between a region gapped by a super-conductor and a region gapped by a ferromagnet carries a "fractional Majorana mode", with a central charge of one half, and a tunneling density of states that vanishes as a high power of the energy \cite{Sagi2014}.

 How shall we identify a 3D fractional topological insulator when we see one?  The strongest signature is probably the surface fractional quantum Hall effect. Note, this is an anomalous quantum Hall effect which emerges as a consequence of breaking of time reversal symmetry, not necessarily by an orbital magnetic field. Other signatures that involve an injection/extraction of electrons into/from the surface, such as STM, photoemission and proximity coupling to super-conductors are expected to be suppressed relative to their counterparts in the un-fractionalized case \cite{Swingle:prb12}. Again, this is in analogy to the fractional quantum Hall effect where the physics of the fractionally charged composite fermions is readily reflected in the probes that do not inject or extract electrons, such as conductivity and sound absorption, and is hard to extract with probes such as tunneling density of states.

 To conclude this section, we note that -- in contrast to the case described in the previous section -- here  the 3D bulk may accommodate and transfer fractionalized excitations between the two surfaces, so each of the surfaces may be in several topological sectors.

 \section{Conclusions}

 Three decades ago the experimental discovery of the fractional quantum Hall effect caught theoretical physicists by utter surprise. Understandably, the lesson has been learned, and much theoretical effort has been devoted to make sure that with fractional topological insulators theory will precede experiment. This effort has employed heavy theoretical machinery, including sophisticated many-body techniques, numerical calculations and above all the deep and profound outcomes of three decades of study.

 In this review we surveyed some of the theoretical understanding of what can and cannot happen in gapped fractionalized electronic states that conserve charge and are symmetric to time reversal, known as fractional topological insulators. We described the physics of 2D and 3D fractional topological insulators, with particular emphasis on their edges and surfaces. We saw that abelian fractional topological insulators may be classified, and examined the stability of their gapless edges to perturbations that are TR symmetric. We surveyed the physics of fractionalized states on the surfaces of 3D un-fractionalized topological insulators. Finally, we reviewed several possible routes to fractional topological insulators in 3D and their expected properties.

 All the above notwithstanding, it seems that the theoretical community would gladly transfer the torch now to the hands of material scientists and experimentalists. There are some directions the torch may be directed to, such as materials that combine strong spin-orbit coupling together with strong interactions (Samarium hexaboride and pyrochlore irridates, for example \cite{witczak-krempa2014,Dzero2013}) or cold atoms systems, but it may very well be that the most promising direction is still in the dark.

%Disclosure
\section*{DISCLOSURE STATEMENT}
The authors is not aware of any conflict of interest relevant to this review.

% Acknowledgements
\section*{ACKNOWLEDGMENTS}
 I am indebted to all my collaborators on subjects related to this review, in particular to Jason Alicea, Erez Berg, Fiona Burnel, David Clarke, Andrew Essin, Paul Fendley, Matthew Fisher, Maciej Koch Janucz, Bert Halperin, Charlie Kane, Michael Levin, Netanel Lindner, Roger Mong, David Mross, Chetan Nayak, Yuval Oreg, Gil Refael, Zohar Ringel, Eran Sela and Kirill Shtengel.

 This research was supported by Microsoft's  Station  Q,  the  European  Research  Councilunder  the  European  Union's  Seventh  Framework  Programme  (FP7/2007-2013)  /  ERC  Project  MUNATOP, the  US-Israel  Binational  Science  Foundation  and  theMinerva Foundation.
% References
%
% Margin notes within bibliography
%\bibnote[<vertical skip value - optional>]{Bibliography margin text.}
%
%\bibliography{ftiannrev}

\begin{thebibliography}{99}
\expandafter\ifx\csname natexlab\endcsname\relax\def\natexlab#1{#1}\fi

\bibitem{FieteMaciejko2015}
Maciejko J, Fiete GA. 2015.
\textit{Nat Phys} 11:385--388

\bibitem{JasonPaul}
Alicea J, Fendley P. 2015.
\textit{Annual Review of Condensed Matter Physics}

\bibitem{BHZ}
Bernevig BA, Hughes TL, Zhang SC. 2006.
\textit{Science} 314:1757

\bibitem{FuTI}
Fu L, Kane CL, Mele EJ. 2007.
\textit{Phys. Rev. Lett.} 98:106803

\bibitem{MooreTI}
Moore JE, Balents L. 2007.
\textit{Phys. Rev. B} 75:121306

\bibitem{RoyTI}
Roy R. 2009.
\textit{Phys. Rev. B} 79:195322

\bibitem{KaneReview}
Hasan MZ, Kane CL. 2010.
\textit{Rev. Mod. Phys.} 82:3045--3067

\bibitem{QiReview}
Qi XL, Zhang SC. 2011.
\textit{Rev. Mod. Phys.} 83:1057--1110

\bibitem{MooreReview}
Moore JE. 2010.
\textit{Nature} 464:194--198

\bibitem{TopologicalFieldTheoryTI}
Qi XL, Hughes TL, Zhang SC. 2008.
\textit{Phys. Rev. B} 78:195424

\bibitem{QHE}
Prange RE, Girvin SM. 1987.
The quantum hall effect.
New York: Springer-Verlag

\bibitem{TKNN}
Thouless DJ, Kohmoto M, Nightingale MP, den Nijs M. 1982.
\textit{Phys. Rev. Lett.} 49:405--408

\bibitem{ThoulessNiu}
Niu Q, Thouless DJ, Wu YS. 1985.
\textit{Phys. Rev. B} 31:3372--3377

\bibitem{MajoranaQSHedge}
Fu L, Kane CL. 2009.
\textit{Phys.\ Rev.\ B} 79:161408(R)

\bibitem{Wen}
Wen XG. 2004.
Quantum field theory of many-body systems.
New York: Oxford

\bibitem{PhysRevB.91.245144}
Sagi E, Oreg Y, Stern A, Halperin BI. 2015.
\textit{Phys. Rev. B} 91:245144



\bibitem{Barkeshliparafendleyons1}
  Barkeshli M. Jian CM and Qi XL 2012. \textit{Phys. Rev. X} 2, 031013,



\bibitem{Barkeshliparafendleyons2}
Barkeshli M, Qi XL. 2013 \textit {Phys. Rev. B} 87, 045130
%{Synthetic Topological Qubits in Conventional Bilayer Quantum {H}all Systems}.
%Unpublished

\bibitem{LindnerParafendleyons}
Lindner NH, Berg E, Refael G, Stern A. 2012.
\textit{Phys. Rev. X} 2:041002

\bibitem{ClarkeParafendleyons}
Clarke DJ, Alicea J, Shtengel K. 2013.
\textit{Nature Commun.} 4:1348

\bibitem{Chengparafendleyons}
Cheng M. 2012.
\textit{Phys. Rev. B} 86:195126


\bibitem{Vaeziparafendleyons}
  Vaezi, A. 2013 \textit {Phys. Rev. B} 87, 035132

\bibitem{Levin:prl09}
Levin M, Stern A. 2009.
\textit{Phys. Rev. Lett.} 103:196803

\bibitem{Bernevig:prl06}
Bernevig BA, Zhang SC. 2006.
\textit{Phys. Rev. Lett.} 96:106802

\bibitem{Young2008}
Young MW, Lee SS, Kallin C. 2008.
\textit{Phys. Rev. B} 78:125316

\bibitem{PhysRevB.90.115426}
Klinovaja J, Tserkovnyak Y. 2014.
\textit{Phys. Rev. B} 90:115426

\bibitem{PhysRevB.90.245401}
Repellin C, Bernevig BA, Regnault N. 2014.
\textit{Phys. Rev. B} 90:245401

\bibitem{PhysRevB.85.195113}
Chen H, Yang K. 2012.
\textit{Phys. Rev. B} 85:195113

\bibitem{Kitaev2003}
Kitaev A. 2003.
\textit{Ann. Phys. (N.Y.)} 303:2--30

\bibitem{PhysRevB.66.205104}
Senthil T, Motrunich O. 2002.
\textit{Phys. Rev. B} 66:205104

\bibitem{PhysRevB.67.115108}
Motrunich OI. 2003.
\textit{Phys. Rev. B} 67:115108

\bibitem{Levin2011}
Levin M, Burnell FJ, Koch-Janusz M, Stern A. 2011{\natexlab{a}}.
\textit{Phys. Rev. B} 84:235145

\bibitem{Ruegg:prl12}
R\"uegg A, Fiete GA. 2012.
\textit{Phys. Rev. Lett.} 108:046401

\bibitem{PhysRevLett.107.126803}
Qi XL. 2011.
\textit{Phys. Rev. Lett.} 107:126803

\bibitem{lu2012}
Lu YM, Ran Y. 2012.
\textit{Phys. Rev. B} 85:165134

\bibitem{simon2015fractional}
Simon SH, Harper F, Read N. 2015.
\textit{arXiv preprint arXiv:1506.08197}

\bibitem{PhysRevLett.108.206804}
B\'eri B, Cooper NR. 2012.
\textit{Phys. Rev. Lett.} 108:206804

\bibitem{Maciejko:prb13}
Maciejko J, R\"uegg A. 2013.
\textit{Phys. Rev. B} 88:241101

\bibitem{PhysRevB.89.075137}
Koch-Janusz M, Ringel Z. 2014.
\textit{Phys. Rev. B} 89:075137

\bibitem{Levin2012}
Levin M, Stern A. 2012.
\textit{Phys. Rev. B} 86:115131

\bibitem{neupert2011b}
Neupert T, Santos L, Ryu S, Chamon C, Mudry C. 2011.
\textit{Phys. Rev. B} 84:165107

\bibitem{Mesaros2013}
Mesaros A, Ran Y. 2013.
\textit{Phys. Rev. B} 87:155115

\bibitem{PhysRevB.87.245120}
Nikoli\ifmmode~\acute{c}\else \'{c}\fi{} P. 2013.
\textit{Phys. Rev. B} 87:245120

\bibitem{Levinedgenosymmetry}
Levin M. 2013.
\textit{Phys. Rev. X} 3:021009

\bibitem{KaneZ2}
Fu L, Kane CL. 2006.
\textit{Phys. Rev. B} 74:195312

\bibitem{PhysRevB.88.245136}
Wang C, Levin M. 2013.
\textit{Phys. Rev. B} 88:245136

\bibitem{Mross2015}
David F. Mross, Andrew Essin, Jason Alicea, Ady Stern, "Anomalous Quasiparticle
  Symmetries and Non-Abelian Defects on Symmetrically Gapped Surfaces of Weak
  Topological Insulators", arXiv:1507.01587

  \bibitem{Cappelli2015}
  Cappelli, Andrea and Randellini, Enrico 2015.
\textit{J. Phys. A}, 48, 105404

\bibitem{Koch:prb13}
Koch-Janusz M, Levin M, Stern A. 2013.
\textit{Phys. Rev. B} 88:115133

%\bibitem{Levin:prb11}
%Levin M, Burnell FJ, Koch-Janusz M, Stern A. 2011{\natexlab{b}}.
%\textit{Phys. Rev. B} 84:235145

\bibitem{BondersonTO}
Bonderson P, Nayak C, Qi XL. 2013.
\textit{Journal of Statistical Mechanics: Theory and Experiment} 2013:P09016

\bibitem{WangTO}
Wang C, Potter AC, Senthil T. 2013.
\textit{Phys. Rev. B} 88:115137

\bibitem{ChenTO}
Chen X, Fidkowski L, Vishwanath A. 2014.
\textit{Phys. Rev. B} 89:165132

\bibitem{MetlitskiTO}
Metlitski MA, Kane CL, Fisher MPA. 2013.
A symmetry-respecting topologically-ordered surface phase of 3d electron
  topological insulators.
\textit{ArXiv} 1306.3286

\bibitem{STO}
Mross DF, Essin A, Alicea J. 2015.
\textit{Phys. Rev. X} 5:011011

\bibitem{AFTIMong}
Mong RSK, Essin AM, Moore JE. 2010.
\textit{Phys. Rev. B} 81:245209

\bibitem{AshvinSenthil}
Vishwanath A, Senthil T. 2013.
\textit{Phys. Rev. X} 3:011016

\bibitem{QiFu2015}
Qi Y, Fu L. 2015.
 arXiv:1505.06201

\bibitem{Maciejko:prl10}
Maciejko J, Qi XL, Karch A, Zhang SC. 2010.
\textit{Phys. Rev. Lett.} 105:246809

\bibitem{Swingle:prb11}
Swingle B, Barkeshli M, McGreevy J, Senthil T. 2011.
\textit{Phys. Rev. B} 83:195139

\bibitem{Maciejko:prb12}
Maciejko J, Qi XL, Karch A, Zhang SC. 2012.
\textit{Phys. Rev. B} 86:235128

\bibitem{Maciejko:prl14}
Maciejko J, Chua V, Fiete GA. 2014.
\textit{Phys. Rev. Lett.} 112:016404

\bibitem{PhysRevX.4.041049}
Geraedts SD, Motrunich OI. 2014.
\textit{Phys. Rev. X} 4:041049

\bibitem{park2010dirac}
Park KS, Han H. 2010.
\textit{Physical Review B} 82:153101

\bibitem{Cho2012dyon}
Cho, Gil Young and Xu, Cenke and Moore, Joel E and Kim, Yong Baek 2012
\textit{New J. Phys.}, 14, 115030


\bibitem{PhysRevB.79.045316}
Bais FA, Slingerland JK. 2009.
\textit{Phys. Rev. B} 79:045316

\bibitem{LinLevin15}
Lin CH, Levin M. 2015.
\textit{Phys. Rev. B} 92:035115

\bibitem{JianQi}
Jian CM, Qi XL. 2014.
\textit{Phys. Rev. X} 4:041043

\bibitem{Neupert2014}
T. Neupert, C. Chamon, C. Mudry, and R. Thomale, ``Wire deconstructionism and
  classification of topological phases'', preprint at
  http://arxiv.org/abs/1403.0953.

\bibitem{Sagi2014}
E. Sagi and Y. Oreg, ``Non-Abelian topological insulators from an array of
  quantum wires'', preprint at http://arxiv.org/abs/1403.1791.

\bibitem{PhysRevB.91.205141}
Santos RA, Huang CW, Gefen Y, Gutman DB. 2015.
\textit{Phys. Rev. B} 91:205141

\bibitem{Mong}
Mong RSK, Clarke DJ, Alicea J, Lindner NH, Fendley P, et~al. 2014.
\textit{Phys. Rev. X} 4:011036

\bibitem{KaneWires}
Kane CL, Mukhopadhyay R, Lubensky TC. 2002.
\textit{Phys. Rev. Lett.} 88:036401

\bibitem{TeoKaneChains}
Teo JCY, Kane CL. 2014.
\textit{Phys. Rev. B} 89:085101

\bibitem{Swingle:prb12}
Swingle B. 2012.
\textit{Phys. Rev. B} 86:245111

\bibitem{witczak-krempa2014}
Witczak-Krempa W, Chen G, Kim YB, Balents L. 2014.
\textit{Annu. Rev. Condens. Matter Phys.} 5:57--82

\bibitem{Dzero2013}
Dzero M, Galitski V. 2013.
\textit{JETP} 117:499--507

\end{thebibliography}
%\bibliographystyle{AnnuRev_CondensedMatter}

\end{document}